# Peril v. Promise: IoT and the Ethical Imaginaries

**Funda Ustek-Spilda***
Virt-EU: Values and Ethics in Innovation for Responsible Technology in Europe
Department of Media and Communications, London School of Economics
f.ustek-spilda@lse.ac.uk

**Alison Powell**
Virt-EU: Values and Ethics in Innovation for Responsible Technology in Europe
Department of Media and Communications, London School of Economics
a.powell@lse.ac.uk

**Irina Shklovski**
Virt-EU: Values and Ethics in Innovation for Responsible Technology in Europe
Business IT
IT University of Copenhagen
irsh@itu.dk

**Sebastián Lehuedé**
Virt-EU: Values and Ethics in Innovation for Responsible Technology in Europe
Department of Media and Communications, London School of Economics
s.lehuede@lse.ac.uk

**ABSTRACT**

The future scenarios often associated with IoT oscillate between the peril of IoT for the future of humanity and the promises for an ever-connected and efficient future. Such a dichotomous positioning creates problems not only for expanding the field of application of the technology, but also ensuring ethical and responsible design and production.

**ACKNOWLEDGMENTS**

This work was supported by the EU Commission Horizon 2020 FP7 Programme, Project no. 732027.

As part of Virt-EU: Values and Ethics in Innovation for Responsible Technology in Europe (EU Horizon 2020), we have conducted ethnographic research into the main hubs of IoT in Europe, such as London, Amsterdam, Barcelona and Belgrade, with developers and designers of IoT to identify the challenges they face in their day-to-day work. In this paper, we focus on the IoT and the ethical imaginaries explore the practical challenges IoT developers face when they are designing, producing and marketing IoT technologies. We argue that top-down ethical frameworks that overlook the *situated* capabilities of developers or the 'solutionist' approaches that treat ethical issues as technical problems are unlikely to provide an alternative to the dichotomous imaginary for the future.

INTRODUCTION

Technology reporting of IoT is characterized by a dichotomous imaginary of the future. On the one hand, it features anxieties about the consequences of pervasive connectivity, on the other hand, it envisions of a future where all technologies will be seamlessly connected to provide the most efficient and productive services.[1] The unpredictability of how IoT technologies will evolve and how socio-technical decision-making processes will change have led to a priori assumptions being made about the impossibility of identifying all ethical issues that might arise from IoT.[2] As a consequence, in the literature, we see that discussions about ethics revolve around a limited focus on security and privacy. Moreover, these issues are understood as technical problems that are technical, so fixable, if only the 'appropriate' solutions are adopted.[3] Issues pertaining to equity, equality and trustability arising from the adoption of IoT on the other hand, are vaguely categorized as 'social ethics'[4] and the underlying ethical, social and economic issues are ignored, so are the situated contexts within which developers and designers of IoT technologies work.

It is against this background, at Virt-EU: Values and Ethics in Innovation for Responsible Technology in Europe (Horizon 2020) we have been studying how developers and designers of IoT technologies approach, discuss and implement ethics in their work. We are interested in understanding how local culture and network society influence the understanding and movement of particular social values among IoT developers, beyond the technical considerations of privacy and security, and the *space* (understood in the Bourdieusian sense) they have for implementing their values into the IoT products they build. We have conducted ethnographical fieldwork into the main IoT hubs in Europe, such as London, Amsterdam, Copenhagen, Barcelona and Belgrade, and also conducted co-design workshops with developers and followed their online discussions about ethics.

In this paper, we present our conceptual framework based on Virtue Ethics, Capability Approach and Care Ethics for understanding and studying socio-cultural, ethical and practical challenges IoT developers face, and illustrative findings from our ethnographical fieldwork to discuss how more care-ful new directions for IoT can be imagined.

A PRACTICAL FRAMEWORK FOR ETHICS

As connected devices and services proliferate, data collection and algorithmic processing become less visible, more ubiquitous and potentially more invasive for users. In such IoT-instrumented spaces, the onus of ethical decision-making about what data ought to be collected, how it should be processed and put in to use shifts further onto those developing and deploying the relevant technologies and services. In conceptualizing ethics as values in action, we draw upon the basic idea that ethics is a process of the application of values in human conduct and this process guides understanding and decision-making. Such a positioning entails that values are not simply individual ideals, but they entail a position of power for those who decide to act or not act on them. It follows then, there might be discrepancies between the values that come to be expressed and values that get enacted in practice. In other words, ethics is values in action and it takes place in contexts – they are *situated*[5] in power relations and constraints.

Such a standpoint permits us to engage with a range of different ways of thinking about ethics. Modern writing on ethical concerns with regard to technology leverages a range of different ethical approaches. By and large, however, these approaches converge towards consequentialist and utilitarian ethics, that is they tend to make their ethical evaluations of their actions based on their consequences. For example, we often see various examples of different versions of the age-old trolley problem discussed in media and in conversations with developers. The question is almost always simplified into an either/or (one person definitely needs to be killed)[6] and the ethical justifiability of killing someone is rarely questioned. Part of the problem with consequentialism is their need to include both potential mundane and existential difficulties, making the rational calculus intractable and often leading to significant reductionism.

We propose to go beyond the consequentialist/utilitarian points of view, by bringing together three ethical frameworks that we think fit better with the problems at hand. These include virtue ethics, capabilities approach and care ethics. Virtue ethics focuses on individual's process of attempting to live a good life[7]; capabilities approach examines their ability to act, including to choose an alternative given the existing structural constraints and opportunities[8]; and care ethics takes into account the shifting obligations and responsibilities of individuals as they are positioned in a web of socio-technical networks.[9] Bringing these three approaches together enables us to acknowledge that ethics as a process is not exclusively dependent on subjectivities of individuals (e.g. their principles and actions), but acknowledges the *situatedness* of ideals and actions within structural conditions that can limit and shape them, and the demands and obligations that arise from these conditions.[10]

IOT AND THE ETHICAL IMAGINARY

Where may agency be located in the digitally mediated world? Mansell, following Taylor, defines social imaginaries as "how things are understood" and "how they come to constitute a moral order which tells us what 'rights and obligations in regard to each other'" (Taylor 2002, 93 cited in Mansell).[11] She isolates three sets of ideas and social imaginaries. The first is the market-led diffusion model whereby technological change in the digital world is considered to be emergent and unpredictable. She notes that any re-distribution of resources, such as information, money and skills, for justice or fairness remains outside the framework of this model, as any intervention in the commercial market is presumed to increase the risk of unpredictable incomes. The second is a state and market-led diffusion model. Here, the social imaginary considers state intervention in the market as essential for the welfare of citizens and for upholding the "rights and obligations we have as individuals in regard to each other" (Ibid, 43). The third and last one is the digital mediation in generative collaborative commons, where civil society and various members of technical communities are ensured through peer-to-peer collaboration I the commons. The main premise is that through non-market participation and good will are generative of individual collective agency in the digital world (Ibid, 44).

In the IoT field, we have also observed three ethical imaginaries in line with Mansell's framework. We refer to these as follows:
1) Technology will sort itself out
2) More regulation is needed
3) Conscious consumers and developers will push for ethical IoT

In the first imaginary, which we called "technology will sort itself out" the main proposition made by IoT developers and designers is that it is because IoT technologies are *still* relatively new and there are many technological aspects that need to be figured out, that the ethical risks associated with it are high. Once the technology matures, it will sort itself out, that is, in the long run, what come to be associated as ethical risks, including privacy and security risks, will no longer constitute problems because there will already be technological solutions for them. This perspective, however, assumes that there will come a point in time where technological development will mature and stall, and there will not be any 'unknown unknowns' left to be discovered. It also runs the risk of assuming it is possible to fix everything through technology, and disregards the problems technology might cause along the way, such as environmental, social and economic issues that we face today.

In the second imaginary "more regulation is needed", more state involvement is considered to be necessary. Some developers argued that because regulation is seen as stalling technological development, IoT companies [and other technology companies in general] will not regulate themselves, unless they have to. The General Data Protection Regulation (GDPR), for instance, is seen as a step forward in this regard, which pushed the companies to pay more attention to issues of privacy. Developers, however, also acknowledge that regulation is unlikely to be able to keep up with the speed of technology, so there will always be a lag between technological risks and regulations that protect individuals.

The third imaginary focuses on conscious consumers and developers who would like to challenge the existing system and create a better one. IoT Manifestos[12], open software, open hardware movements, trustmark, trademark and other interventions for ethical certification of IoT technologies as well the demand from consumers for environmentally, socially and economically 'conscious' products are considered as examples. Nevertheless, the *capabilities* of developers working in sub-contracted positions or un-established start-ups to challenge the commercial infrastructure they are part of is also mentioned as a major challenge for the third imaginary to shape the future of IoT. This is why, many developers quickly revert to mentioning 'consumers' and how their choices might shape the direction of IoT. If they refuse buying products that are not ethically-designed and built, and that, there is a 'real' market interest for building ethical products, then the approach of working with 'beta-versions' of products until a major privacy or security breach (or other ethical risks) might change. This line of thinking, however, continues the thread that is apparent in the previous two imaginaries: postponing the ethical decision-making now, and shifting the responsibility to others.

DISCUSSION AND EXPECTATIONS FROM THE WORKSHOP

In this paper, we presented a brief introduction to our ongoing research at Virt-EU Project and discussed the three ethical imaginaries that we have observed in our field research with designers and developers of IoT products. We argue that top-down ethical frameworks, including regulations, that overlook the *situated* capabilities of developers or the 'solutionist' approaches which treat ethical issues as technical problems are unlikely to provide an alternative to the dichotomous directions drawn for IoT.

In this workshop, against this background, we would like to discuss our conceptual framework: Virtue Ethics, Capability Approach and Care ethics for not only studying ethical challenges the IoT raises today, but also the future, including autonomous systems, collaborative aspects and sensitive technologies such as in the context of wearables in the healthcare and social care sectors.